\documentclass[floatfix,aps,nofootinbib,showpacs]{revtex4}
\begin{document}

\rightline{April 2003}

\title{Was ordinary matter synthesised from mirror matter?\\
An attempt to explain why
$\Omega_{Baryon} \approx 0.2\Omega_{Dark}$.}

\author{R. Foot}\email{foot@physics.unimelb.edu.au}
\author{R. R. Volkas}\email{r.volkas@physics.unimelb.edu.au}
\affiliation{School of Physics, Research Centre for High Energy Physics,
The University of Melbourne, Victoria 3010, Australia}

\begin{abstract}
The cosmological dust has begun to settle. A likely picture
is a universe comprised (predominantly) of three components:
ordinary baryons ($\Omega_B \approx 0.05$), non-baryonic 
dark matter ($\Omega_{Dark} \approx 0.22$) and dark energy
($\Omega_{\Lambda} \approx 0.7$). We suggest that the
observed similarity of the abundances of ordinary baryons
and non-baryonic dark matter ($\Omega_{B}/\Omega_{Dark} \approx 0.20$)
hints at an underlying similarity between the fundamental properties of
ordinary and dark matter particles.
This is necessarily the case if dark matter is 
identified with mirror matter. We examine a specific 
mirror matter scenario where $\Omega_B/\Omega_{Dark} \approx 0.20$
is naturally obtained.
\end{abstract}
\pacs{95.35.+d,98.80.Cq,11.30.Er,12.90.+b}

\maketitle

Data obtained from high redshift supernovae \cite{hr}, cosmic
microwave background anisotropy measurements (culminating
with the recent WMAP results \cite{cmb}), and other sources,
have greatly enriched cosmology. These data
are consistent with a spatially-flat universe ($\Omega_{tot} \simeq 1.0$), 
composed (predominantly) of ordinary matter ($\Omega_B \approx 0.05$),
non-baryonic dark matter ($\Omega_{Dark} \approx 0.22$) and 
dark energy ($\Omega_{\Lambda} \approx 0.7$), although an alternative
interpretation of the data with a zero $\Omega_{\Lambda}$ has
also been proposed \cite{alternative}.

One striking feature of these results
is the fact that the dark energy is currently of the same order
of magnitude as the present mass density, $\Omega_m = \Omega_B + \Omega_{Dark}$,
of the universe.
Because dark energy and matter densities scale differently with time, 
it is expected
that this feature was not true at earlier epochs.
The present day similarity of dark energy and matter
densities has been called the cosmic coincidence problem.

A fascinating related feature, which has not been emphasised so much 
in the literature,
is the similarity in magnitude of the ordinary and dark matter
densities:
\begin{equation}
\Omega_{B}/\Omega_{Dark} \approx 0.20.
\end{equation}
In contrast to $\Omega_{m}/\Omega_{\Lambda}$, the
ratio of dark matter to ordinary matter is expected to be constant in time
until a very early epoch.
This means that the amount of dark matter produced
in the early universe is of the same order of magnitude
as the ordinary matter, despite their obviously
disparate properties.

The observed similarity in the abundances of
ordinary and dark matter   
hints at an underlying similarity between
the microscopic properties
of the elementary particles comprising the ordinary matter and the dark matter.
Clearly, the standard exotic weakly interacting dark matter
scenarios seem to offer no hope in explaining this cosmic
coincidence because these particles have completely
different properties (different masses and interactions)
from the ordinary baryons. A priori, a dark matter/ordinary
matter ratio of, say, $10^6$ would appear to be equally likely in 
these scenarios.

Interestingly, there is one obvious candidate for the dark matter
which does actually require a similarity between the properties of 
ordinary and dark matter.
This is ``mirror matter'', a sector whose existence is required if nature
respects a fundamental exact parity (mirror) symmetry \cite{parity,parity2}.
In this scenario, each ordinary particle has a 
distinct `mirror partner' {\it of the same mass}
connected with the ordinary particles via the parity
symmetry:
\begin{eqnarray}
&x \to -x,\quad t \to t, \nonumber \\
&W^{\mu} \leftrightarrow W'_{\mu}, \quad
B^{\mu} \leftrightarrow B'_{\mu}, \quad G^{\mu} \leftrightarrow G'_{\mu}
\nonumber \\
&\ell_{iL} \leftrightarrow \gamma_0 {\ell'}_{iR},\ \  
e_{iR} \leftrightarrow \gamma_0 {e'}_{iL},\ \ 
q_{iL} \leftrightarrow \gamma_0 {q'}_{iR},\ \ 
u_{iR} \leftrightarrow \gamma_0 {u'}_{iL},\ \ 
d_{iR} \leftrightarrow \gamma_0 {d'}_{iL},
\end{eqnarray}
where $G^{\mu}, W^{\mu}, B^{\mu}$
are the standard $G_{SM} \equiv SU(3)\otimes SU(2)\otimes U(1)$ gauge particles,
$\ell_{iL}, e_{iR}, q_{iL}, u_{iR}, d_{iR}$ are the standard leptons and quarks
($i=1,2,3$ is the generation index) and the primes denote
the mirror particles.
There is also a standard Higgs doublet $\phi$ with a mirror Higgs doublet 
partner $\phi'$,
and it can be shown that $\langle \phi \rangle = \langle \phi' \rangle$
for a large range of parameters of the Higgs potential. We adopt
this parameter regime, making the mirror symmetry 
exact (i.e.\ not spontaneously broken) \cite{parity2}.

Clearly, in this theory the properties of the mirror particles
exactly mirror
those of the ordinary ones. In particular, the mirror baryons are
stable and do not couple to ordinary photons so they {\it necessarily}
have the right broad properties to be the non-baryonic dark matter.
Furthermore, recent detailed studies of large scale structure formation
have confirmed that mirror baryons are a viable dark matter
candidate from that point of view \cite{detail}.
Thus, if mirror symmetry is a fundamental symmetry of nature,
it is natural to set $\Omega_{Dark} = \Omega'_{B}$, the
mirror baryon mass density.
Of course, if dark matter is identified as mirror matter then
mirror stars, mirror planets and even mirror space-bodies
should exist; there is actually fascinating evidence for these
things (for reviews and references, see Ref.\cite{review}). 

In this interpretation one would expect $\Omega_B = \Omega'_{B}$
if the initial conditions of the universe were also mirror symmetric
and no macroscopic asymmetry (such as a temperature difference) 
was produced during the early evolution of the universe.
However, the success of standard big bang nucleosynthesis (BBN) 
does suggest
that $T'$ was somewhat less than $T$ during the BBN epoch,
\begin{eqnarray}
T'/T \stackrel{<}{\sim} 0.5 \ \ {\rm at} \ \ T \sim 1\ MeV,
\end{eqnarray}
in order for the expansion of the universe to have been within
an acceptable range.
If the temperatures are different, then this means
that either the initial conditions of the universe
were asymmetric or that the asymmetry was
induced during the evolution of the universe,
due to, for example, an asymmetric fluctuation that
became enhanced in some way.
One such way is via inflation. One can imagine having an
`ordinary inflaton' coupling to ordinary matter,
and a `mirror inflaton' coupling to mirror matter \cite{inflation}.
If inflation is triggered by some random fluctuation, then
it can occur in the two sectors at different times, leading
to $T \neq T'$ after reheating in the two sectors.
In such a scenario, one expects the baryon number and 
mirror baryon number to be unequal (since baryogenesis
or leptogenesis depends on the temperature and expansion rate).

Provided that the temperatures
of the two sectors are not too different  
this might explain the fact that $\Omega_B$ is within an
order of magnitude of $\Omega'_{B}$. 
Clearly, the details will depend on the precise model for
baryogenesis used by nature, which is of course not known
(see the first paper of Ref.\cite{detail} for a couple of examples).

We consider instead an alternative possibility: that immediately 
after reheating,
either $\Omega'_B \gg \Omega_B$ or the opposite obtains, perhaps
due to a very large initial temperature hierarchy (e.g.\ $T' \gg T$).  
Instead of specifying models in great detail, we will try to
explain the ratio $\Omega_{B}/\Omega'_{B} \approx 0.2$
in an almost model independent way
using sphaleron \cite{krs} and other processes to transfer 
asymmetries. 

In particular, consider the following scenario:
\begin{itemize}
\item
{\it Step 1}.\
A period of inflation sufficiently long to solve
the standard cosmological problems (flatness, homogeneity and so on). 
If this period of inflation is driven by the vacuum energy of a `mirror
inflaton', then after reheating one is
left with a large temperature asymmetry, $T' \gg T$.  

\item
{\it Step 2}.\
At a certain temperature, $T' = T_1$, 
an asymmetry generation process
takes place: the out-of-equilibrium
decay of a heavy mirror lepton \cite{fuk}, for example.
This generates initial $B'$ and/or $L'$ asymmetries.
No significant $B$ or $L$ will be generated if the
temperature $T$ of the ordinary sector is sufficiently low.

\item
{\it Step 3}.\
Ordinary baryon number (and lepton number) will
be generated from $B'$ or $L'$ if there is some process
that can transfer asymmetries from the mirror sector
to the ordinary sector. 
The simplest interaction that can do this 
(involving standard model particles and mirror particles) is the 
non-renormalisable dimension-5 operator,
\begin{equation}
L = {1 \over M_N} \bar \ell_L \phi^c \ell'_R \phi' + H.c.,
\label{nr}
\end{equation}
where $M_N$ is a large scale.
Such an operator can be related to the neutrino masses,
$m_{\nu} \sim \langle \phi \rangle^2/M_N$,
and can be motivated by neutrino physics experiments.
This interaction will bring the ordinary and mirror sectors
into thermal equilibrium
for temperatures $M_N \stackrel{>}{\sim} 
T \stackrel{>}{\sim} 10^{10} (eV/m_{\nu})^2$ GeV, 
combining with other interactions (such as those induced by sphalerons)
to generate $B$ and $L$ from the initial $B'$ and/or $L'$ asymmetries produced
in step 2.

\item
{\it Step 4}.\
We require a second, but relatively short period of inflation
(or some other type of mechanism)
to induce the mild hierarchy $T'/T \stackrel{<}{\sim} 0.5$ 
suggested by BBN. It is tempting to associate the
second period of inflation with the existence
of an ordinary inflaton [perhaps the mirror partner of
the mirror inflaton of step(i)]. In addition to generating
the temperature asymmetry ($T'/T \stackrel{<}{\sim} 0.5$), the
reheating associated with the ordinary inflaton will dilute
both the ordinary and mirror baryon numbers relative to the entropy;
the amount of reheating after the second period of inflation must
therefore be moderate, consistent with the quite mild temperature
asymmetry required for consistency with BBN.\footnote{Note, however,
that neither inflation nor the reheating processes directly
affect the final value of the ratio $\Omega_B/\Omega'_B$. To 
see this note that the (current) value of the ratio,
$\Omega_B/\Omega'_B$ can be related to the absolute numbers
of baryons ($N_B$) and mirror baryons ($N'_B$) as follows:
$\Omega_B/\Omega'_B = (N_B m_B/V/\rho_c)/(N'_B m_B/V/\rho_c)
= N_B/N'_B$
where $V$ is the characteristic volume and 
$\rho_c$ is the critical density. 
Since the absolute numbers of baryons and mirror baryons are
not affected by inflation or reheating, it follows that
the ratio
$\Omega_B/\Omega'_B$ is also not directly affected by either of these processes.}

\end{itemize}

\noindent
It turns out that this scenario allows for a definite
prediction of the ratio $\Omega_B/\Omega'_B$, typically about $0.2$
as we will show, consistent with the observations. 

For definiteness, we will assume the existence of only one
operator of the form of Eq.(\ref{nr}), coupling to the second
generation,
\begin{eqnarray}
L = {1 \over M_N} \bar \ell_{2L} \phi^c \ell'_{2R} \phi' + H.c.
\label{nr2}
\end{eqnarray}
We expect that similar results would arise if the 
interaction, Eq.(\ref{nr2}),
was replaced by some other type of interaction, so long as
it coupled ordinary to mirror particles. The use of the
specific form of Eq.(\ref{nr2}) is only meant to provide a 
definite illustration of our idea.

In order to proceed, we need to know the
temperature above which the interaction, Eq.(\ref{nr2}),
is efficient enough to bring the ordinary sector
into equilibrium with the mirror sector.
We expect this to be quite high,
$T \stackrel{>}{\sim} T_2 \sim 10^{10}$ GeV (for $m_{\nu}
\stackrel{<}{\sim} 1$ eV).
For temperatures less than about $10^{12}$ GeV, the
QCD and electroweak sphaleron processes plus 
the Yukawa interactions for the
$c, t, b, \tau$ fermions \cite{reviewb} are faster than the
expansion rate of the universe. 
However, until the temperature drops below about 
$10^{10}$ GeV, Yukawa interactions
are not strong enough to keep $e_R$, $\mu_R$, $s_R$, $d_R$ and $u_R$
(and their mirror counterparts) in chemical equilibrium and we can
neglect these processes for now. 
Thus, for a temperature $T 
\sim 10^{10}$ GeV, we have the following 
chemical potential 
constraints:\footnote{Note that the $SU(2)$ gauge interactions ensure that
$\mu_{q_{1L}} = \mu_{q_{2L}} = \mu_{q_{3L}} \equiv \mu_{q_{L}}$
(due to processes such as $\bar u_{3L} d_{3L} \rightarrow 
\bar u_{3L} d_{2L}$, \ $\bar u_{2L} d_{2L} \rightarrow \bar u_{2L} d_{1L}$).
Also, to simplify notation, we have used the abbreviations $\mu_{u_{i}}
\equiv \mu_{u_{iR}}$, $\mu_{d_{i}} \equiv \mu_{d_{iR}}$, $\mu_q \equiv \mu_{q_{L}}$,
and $\mu_{\ell_{i}} \equiv \mu_{\ell_{iL}}$.}
\begin{eqnarray}
9\mu_q &+& \sum_{i=1}^3 \mu_{\ell_i} = 0
\ \ \ [{\rm Electroweak\ sphaleron}],
\nonumber \\
6\mu_q &-& \sum_{i=1}^3 (\mu_{u_i} + \mu_{d_i}) = 0 \ \ \ [{\rm QCD \ sphaleron}],
\nonumber \\
3\mu_q &+& 2\mu_{\phi} + \sum_{i=1}^3 (2\mu_{u_i} - \mu_{d_i} - \mu_{\ell_i}  -
\mu_{e_i}) 
 = 0 \ \ \ [{\rm Hypercharge \ neutrality}],
\nonumber \\
\mu_q &-& \mu_{\phi} - \mu_{d_3} = 0 \ \ \ [{\rm Yukawa\ interactions}], 
\nonumber \\
\mu_q &+& \mu_{\phi} - \mu_{u_2} = 0 \ \ \ [{\rm Yukawa\ interactions}],
\nonumber \\
\mu_q &+& \mu_{\phi} - \mu_{u_3} = 0 \ \ \ [{\rm Yukawa\ interactions}],
\nonumber \\
\mu_{\ell_3} &-& \mu_{\phi} - \mu_{e_3} = 0 \ \ \ [{\rm Yukawa\ interactions}].
\label{mon}
\end{eqnarray}
For each of the above equations there is
a mirror equation (for the mirror particles)
which is of the same
form except with primes on all the chemical potentials.
Note that the dimension-5 operator, Eq.(\ref{nr2}),
couples the ordinary and mirror particles together, 
leading to one more equation:
\begin{eqnarray}
-\mu_{\ell_2} - \mu_{\phi} + \mu'_{\ell_2} + \mu'_{\phi} = 0.
\label{mon2}
\end{eqnarray}
Thus, we have a total of fifteen equations constraining twenty-eight chemical
potentials. This leaves thirteen independent linear combinations
of chemical potentials, correlated with
thirteen easily identified conserved charges.

There are six conserved 
charges in the ordinary sector and six in the mirror sector,
\begin{eqnarray}
{\cal L}_1 &= {1 \over 3} B - L_1,\ \ 
&{\cal L}_1' = {1 \over 3} B' - L_1'
\nonumber \\
{\cal L}_2 &= {1 \over 3}B - L_3,\ \ 
&{\cal L}_2' = {1 \over 3}B' - L_3'
\nonumber \\
{\cal L}_3 &= L_{e_{1R}}, \ \ \ \ 
&{\cal L}_3' = L'_{e_{1R}}
\nonumber \\
{\cal L}_4 &=  L_{e_{2R}}, \ \ \ \
&{\cal L}_4' = L'_{e_{2R}}
\nonumber \\
{\cal L}_5 &= B_{u_{1R}} - B_{d_{1R}}, \ \ 
&{\cal L}_5' = B'_{u_{1R}} - B'_{d_{1R}}
\nonumber \\
{\cal L}_6 &= B_{d_{1R}} - B_{d_{2R}},\ \ 
&{\cal L}_6' = B'_{d_{1R}} - B'_{d_{2R}}
\label{mon3}
\end{eqnarray}
where $L_{1,2,3}$ are family lepton numbers,
$L_{e_{1R}}$ is the right-handed electron number
(which has a value of one for $e_{1R}$ and zero for all other fields)
and with the other charges defined similarly. To identify
the thirteenth conserved charge, observe that
the interaction of Eq.(\ref{nr2}) implies that only the sum of
${1 \over 3}B - L_2$ and
${1 \over 3}B' - L_2'$,
\begin{eqnarray}
{\cal L}_0 = {1 \over 3}B - L_2 + {1 \over 3}B' - L_2',
\label{mon5}
\end{eqnarray}
is conserved. It is straightforward to write down the 
chemical potential linear combinations that correspond
to each of these charges, using, for instance,
\begin{eqnarray}
B & \leftrightarrow & 6 \mu_q + \sum_{i=1}^3 (2 \mu_{u_{i}} + \mu_{d_{i}}),
\nonumber\\
L_1 & \leftrightarrow & 2 \mu_{\ell_{1}} + \mu_{e_{1}},
\end{eqnarray}
and so on. 

Equations (\ref{mon}) and (\ref{mon2}) can be solved to obtain the
baryon and lepton numbers (after chemical reprocessing), in terms of
the conserved quantities ${\cal L}_i$, ${\cal L}_i'$ and ${\cal L}_0$:
\begin{eqnarray}
B &=&  \alpha_0 {\cal L}_0 +
\sum_{i=1}^6 \alpha_i {\cal L}_i + \sum_{i=1}^6 \alpha'_i {\cal L'}_i,
\nonumber \\
L &=& \beta_0 {\cal L}_0 + \sum_{i=1}^6 \beta_i {\cal L}_i + 
\sum_{i=1}^6 \beta'_i {\cal L'}_i .
\label{wed}
\end{eqnarray}
Under mirror symmetry, $B \leftrightarrow B'$, $L
\leftrightarrow L'$, ${\cal L}_i \leftrightarrow {\cal L}'_i$
(and ${\cal L}_0 \to {\cal L}_0$). Hence,
\begin{eqnarray}
B' &=& \alpha_0 {\cal L}_0 + \sum_{i=1}^6 \alpha_i {\cal L'}_i 
+\sum_{i=1}^6 \alpha_i' {\cal L}_i,
\nonumber \\ 
L' &=& \beta_0 {\cal L}_0 + \sum_{i=1}^6 \beta_i {\cal L'}_i 
+ \sum_{i=1}^6 \beta_i' {\cal L}_i.
\label{wed2}
\end{eqnarray}
The values of the $\alpha$ and $\beta$ parameters
are given in Table I.


\vskip 1cm
\begin{center}
{\bf Table I} \, 
The $\alpha$ and $\beta$ coefficients
in the expansions of $B$ and $L$ defined in Eqs.(\ref{wed}) and (\ref{wed2}).
\end{center}
\begin{center}
\begin{tabular*}
{0.75\textwidth}{@{\extracolsep{\fill}}
|c|c|c|c|}  \hline
& & & \\
$\alpha_0 = \frac{69}{316}$ &   &
$\beta_0 = \frac{-89}{316}$ &  \\
& & &
\\
$\alpha_1 = \frac{43263}{98276}$ & $\alpha'_1 = \frac{-345}{98276}$ &
$\beta_1 = \frac{-55803}{98276}$ & $\beta'_1 = \frac{445}{98276}$ \\
& & &
\\
$\alpha_2 = \frac{7050}{24569}$ & $\alpha'_2 = \frac{414}{24569}$ &
$\beta_2 = \frac{-16571}{24569}$ & $\beta'_2 = \frac{-534}{24569}$ \\
& & &
\\
$\alpha_3 = \frac{45189}{98276}$ & $\alpha'_3 = \frac{-6003}{98276}$ &
$\beta_3 = \frac{31443}{98276}$ & $\beta'_3 = \frac{7743}{98276}$ \\
& & &
\\
$\alpha_4 = \frac{23385}{98276}$ & $\alpha'_4 = \frac{15801}{98276}$ &
$\beta_4 = \frac{59567}{98276}$ & $\beta'_4 = \frac{-20381}{98276}$ \\
& & &
\\
$\alpha_5 = \frac{-963}{24569}$ & $\alpha'_5 = \frac{2829}{24569}$ &
$\beta_5 = \frac{5515}{24569}$ & $\beta'_5 = \frac{-3649}{24569}$ \\
& & &
\\
$\alpha_6 = \frac{-963}{49138}$ & $\alpha'_6 = \frac{2829}{49138}$ &
$\beta_6 = \frac{5515}{49138}$ & $\beta'_6 = \frac{-3649}{49138}$ \\
& & &
\\
\hline
\end{tabular*}
\end{center}
\vskip 0.8cm
\noindent

Of course the values of the conserved charges,
${\cal L}_i$, ${\cal L}'_i$ and ${\cal L}_0$ 
depend on the {\it initial} asymmetry generation mechanism.
We consider, for definiteness, the simple case of
non-zero $B'$ and/or $L'$:
$B' = X_0'$, $L' = Y_0'$, $B = L = 0$ (with $L'_{\ell_{1}} =
L'_{\ell_{2}} = L'_{\ell_{3}} \equiv Y_0'/3$ and $B'_{u_{1R}} = B'_{d_{1R}} =
B'_{d_{2R}}$).
In this case the only non-zero conserved charges are
${\cal L}'_1$, ${\cal L}'_2$ and ${\cal L}_0$ with
${\cal L}'_1 = {\cal L}'_2 = {\cal L}_0 = \frac{1}{3}(X_0' - Y_0')
\equiv Z$.
After chemical processing, with Eqs.(\ref{mon}) and (\ref{mon2})
summarising the consequences,
we are left with the baryon and mirror baryon asymmetries 
given by
\begin{eqnarray}
B = Z (\alpha_0 + \alpha'_1  + \alpha'_2),
\nonumber \\
B' = Z (\alpha_0 + \alpha_1 + \alpha_2),
\end{eqnarray}
where Eqs.(\ref{wed}) and (\ref{wed2}) have been used.
So, the ordinary matter/dark matter ratio is given by
\begin{eqnarray}
{B \over B'} = {\alpha_0 + \alpha'_1  + \alpha'_2 \over \alpha_0 + \alpha_1 +
\alpha_2} \simeq 0.24.
\end{eqnarray}

This is the ratio of ordinary baryon-number density to mirror
baryon-number density at $T \sim 10^{10}$ GeV.
The value of this ratio changes somewhat at lower temperatures as different 
species come into chemical equilibrium. 
However, at temperatures near
the electroweak phase transition, $T = T_{EW} \sim 200$ GeV,
the `final' values of $B$ and $B'$ depend only on 
the values of $B-L$ and $B'-L'$.
These charges are separately conserved for $T \ll 10^{10}$ GeV,
because the interaction in Eq.(\ref{nr2}) is slower than the expansion rate.
This yields the well-known relation between $B$ and $B-L$ \cite{shap}, 
\footnote{Higher order corrections exist, but are quite small
($\stackrel{<}{\sim}$
few percent) and depend on whether the electroweak phase
transition is first order or second order \cite{shap2}.}
\begin{eqnarray}
B = {28 \over 79} (B - L),
\end{eqnarray}
with an identical relation for $B'$ in terms of $B'-L'$ (equally
well-known to mirror physicists).
Thus the final value of the ratio $B/B'$ ,
denoted $B^f/B'^f$ below,
is equal to the ratio $(B-L)/(B'-L')$,
that is,
\begin{eqnarray}
{B^f \over B'^{f}} = {\alpha_0 + \alpha'_1 + \alpha'_2 - \beta_0 - \beta'_1  
- \beta'_2
\over \alpha_0 + \alpha_1 + \alpha_2 - \beta_0 - \beta_1 - \beta_2}
\simeq 0.21.
\end{eqnarray}

Of course, if there is some brief period of inflation between
$T \sim 10^{10}$ GeV and $T_{EW}$
then the above result is only valid provided that the reheating 
temperature of $T'$ and $T$ are both greater than $T_{EW}$.
In the alternative case where
the reheating temperatures satisfy
$T' < T_{EW}$ and $T > T_{EW}$,
then the final asymmetry ratio is
\begin{eqnarray}
{B^f \over B'^f} = {28 \over 79}{\alpha_0 + \alpha'_1 + \alpha'_2 - 
\beta_0 - \beta'_1 - \beta'_2
\over \alpha_0 + \alpha_1 + \alpha_2}
\simeq 0.20,
\end{eqnarray}
which is not very different from the previous result.

While these quantitative results are quite
impressive when compared to the observations,
${\Omega_{b}/\Omega_{dark}} = 0.20 \pm 0.02$ 
according to Ref.\cite{cmb},
one should be cautious.
It is probably unlikely that the standard model description
of the ordinary particle interactions remains
valid up to $T \sim 10^{10}$ GeV: new particles
and interactions may exist. The interaction of Eq.(\ref{nr2}) may
not occur, or if it does it might not be strong enough 
to thermalise the ordinary and mirror sectors.
Some other interaction might be responsible for this,
perhaps involving other exotic heavy fermions or scalars.

On the positive side, it is quite easy to see why 
we obtained an $\Omega_b/\Omega_{b'}$ 
value of about the right size -- given our
assumption that $B$ and $L$ are generated from $B'$ and $L'$.
The point is that this creates a fundamental asymmetry:
there are more non-zero mirror charges
than ordinary ones. In the particular scenario
studied, the only non-zero conserved charges were
${\cal L}'_{1,2}$ and ${\cal L}_0$. These  
are mainly mirror charges, with only
${\cal L}_0$ involving ordinary particles.
Thus, the observation that ordinary matter is
a small, yet significant fraction of the
total matter density in the universe seems to be
explicable, at least in principle, if mirror matter
is the dark matter.

\vspace{1cm}

\acknowledgments{This work was supported by the Australian Research Council.}

\end{document}